\documentstyle[prd,aps,preprint]{revtex}
\tightenlines
\input epsf.tex
\def\DESepsf(#1 width #2){\epsfxsize=#2 \epsfbox{#1}}
\begin{document}

\draft
%@twocolumnfalse\endcsname
\preprint{\hbox{CTP-TAMU-28-99}}
\title{Grand Unification Scale CP Violating Phases And The  Electric Dipole
Moment} 

\author{ E. Accomando, R. Arnowitt and B. Dutta }

\address{ Center For Theoretical Physics, Department of Physics, Texas A$\&$M
University, College Station TX 77843-4242}
\date{July, 1999}
\maketitle
\begin{abstract} The question of CP violating phases in supersymmetry and
electric dipole moments (EDMs) is considered within the framework of
supergravity grand unification (GUT) models with a light ($\stackrel{<}{\sim}$1
TeV) mass spectrum. In the minimal model, the nearness of the t-quark Landau
pole automatically suppresses the t-quark cubic soft breaking phase at the
electroweak scale. However, current EDM data require the quadratic soft breaking
phase to be small at the electroweak scale unless tan$\beta$ is small
(tan$\beta\stackrel{<}{\sim}$3), and the EDM data combined with the requirement
of electroweak symmetry breaking require this phase to be both large and highly
fine tuned at the GUT scale unless tan$\beta$ is small. Non minimal models are
also examined, and generally show the same behavior.
 \end{abstract}

\section{Introduction}

The Standard Model (SM) of strong and electroweak interactions possesses only
one source of CP violating phases: the phase in the CKM matrix. While the
physical origin of this phase remains unknown, it appears to satisfactorily
account for the observed CP violation in the K meson system, and future data
from B factories and high energy accelerators will be able to test its validity
for B meson phenomena. It was early on realized that supersymmetric (SUSY)
extensions of the Standard Model allowed for an array of new CP violating
phases, and that these phases could give large contributions to the electric
dipole moments [EDMs] of the electron and neutron \cite{redm1}, thus violating
the known experimental bounds
\cite{de,dn}. Several resolutions to this problem have been proposed,  e.g. one
might assume the phases are quite small i.e.
$\stackrel{<}{\sim}O(10^{-2})$\cite{redm1,redm3}, or suppress the diagrams by
assuming the relevant SUSY particles are very heavy \cite{redm4}. Both these
suggestions a priori appear unsatisfactory, i.e. the first possibility would
appear to require a significant amount of fine tuning, and the second would
imply a SUSY spectrum in the TeV domain, possibly beyond the reach of even the
LHC. More recently, it has been realized that some cancellations can occur in
different contributions to the EDMs in certain parts of the SUSY parameter 
space \cite{nath1}
which has led to considerable examination of this possibility 
\cite{nath1,nath2,nath3,falk1,bk,falk2,kane1,kane2,bartl,pokorski}. In the
minimal supersymmetric extension of the SM, the MSSM, (which we define as the
low energy SUSY extension of the SM obtained by supersymmetrizing the particle
spectrum with two Higgs doublets) this does appear to reduce the amount of fine
tuning needed, allowing for larger phases. The existence of such phases could
also have effects on other predictions of the MSSM
\cite{demir,carm,falk3,falk4,nath4,pila},
thus allowing future experimental checks of this idea.

In this paper we consider these questions from the viewpoint of supergravity
(SUGRA) grand unified models (GUTs). In particular we consider gravity mediated
models with R-parity invariance, where supersymmetry is broken in a hidden
sector at some scale above the GUT scale $M_G$ (presumably at the string or
Planck scale $M_{Pl}$), the breaking being transmitted to the physical sector by
gravity \cite{acn}. (For previous work from this viewpoint see
Refs.\cite{nath1,nath2,falk1,bk,kane2,bartl}.) Such models are much
more constrained than the MSSM. Thus low energy properties are now determined by
running the renormalization group equations (RGE) from $M_G$ to the electroweak
scale. CP violating phases that in the MSSM are arbitrary parameters now get
correlated by the RGE, and no longer can be assigned independently to satisfy
the EDM experimental bounds. Further, the radiative breaking of $SU(2)\times
U(1)$ puts additional constraints on the CP violating phases. We find that as a
consequence of these constraints, one must still fine tune some of the CP
violating phases at the electroweak scale to be quite small, except for small
regions of the parameter space with tan$\beta$ close to its minimum value, or
when SUSY masses are very  large. Further, except for the lowest tan$\beta$ a
serious fine tuning develops at the GUT scale. These phenomena appear to be true
both for the minimal mSUGRA model with universal soft breaking masses, and with
nonuniversal soft breaking extensions. Hence the required smallness of the
phases also tends to reduce the possibility of collateral evidence in other
phenomena for their existence in SUGRA models.

In Sec.2, we discuss the solutions of the RGE for mSUGRA models and explain
which phases that are large at the GUT scale get naturally suppressed at the
electroweak scale (and which do not), and what the constraints of electroweak
breaking imply both at the tree and 1-loop level. We then examine the amount of
fine tuning needed to satisfy the existing EDM data, and what would be needed if
those data were improved by a factor of 10 without finding any EDM. Sec.3
considers some effects of nonuniversal soft breaking on the EDM results.

In gravity mediated SUGRA models, the structure of the soft breaking parameters
is to be deduced from the nature of the Kahler potential and gauge kinetic
function at the GUT scale \cite{acn,cfgv,wb}. Thus what might be reasonable
sizes for CP violating phases are presumably Planck scale physics questions.
While at present there is no theory of such phenomena, one can still examine the
general framework to see under what circumstances small or large phases might
occur ``naturally'' at the GUT scale. This is analyzed in Sec.4, and a model
with ``naturally'' small CP violating phases is discussed.

Concluding remarks are given in Sec.5. 

\section{EDM For mSUGRA Models} Supergravity GUT models with universal soft
breaking of supersymmetry, mSUGRA, depend upon five  parameters at the GUT
scale: $m_{1/2}$ (the universal gaugino mass), $A_0$ (the cubic soft breaking
mass, $B_0$ (the quadratic soft breaking mass), $\mu_0$ (the Higgs mixing
parameter) and $m_0$ (the universal squark and slepton mass). Of these, the
first four may be complex. However, it is always possible to make a phase
transformation on the gaugino fields to make $m_{1/2}$ real, and since the
reality is preserved by the RGE at 1-loop order, we will assume for now on that
$m_{1/2}$ is real. We parameterize the remaining phases at $M_G$ by :
\begin{eqnarray} A_0&=&|A_0|e^{i \alpha_{0A}};\, B_0=|B_0|e^{i
\theta_{0B}};\,\mu_0=|\mu_0|e^{i
\theta_{0\mu}}.
\label{amb}
\end{eqnarray} The RGE allow one to determine the low energy parameters in terms
of those of Eq.(\ref{amb}) and $m_{1/2}$ and $m_0$. Thus at the electroweak
scale, one obtains a different A parameter for each fermion ($A_u$, $A_d$,
$A_t$, $A_b$,
$A_e$ and
$A_\tau$), real gaugino masses $\tilde m_3$, $\tilde m_2$ and $\tilde m_1$ for
the $SU(3)_C$, $SU(2)_L$ and $U(1)_Y$ sectors, and complex $B$ and $\mu$
parameters.

\begin{figure}[htb]
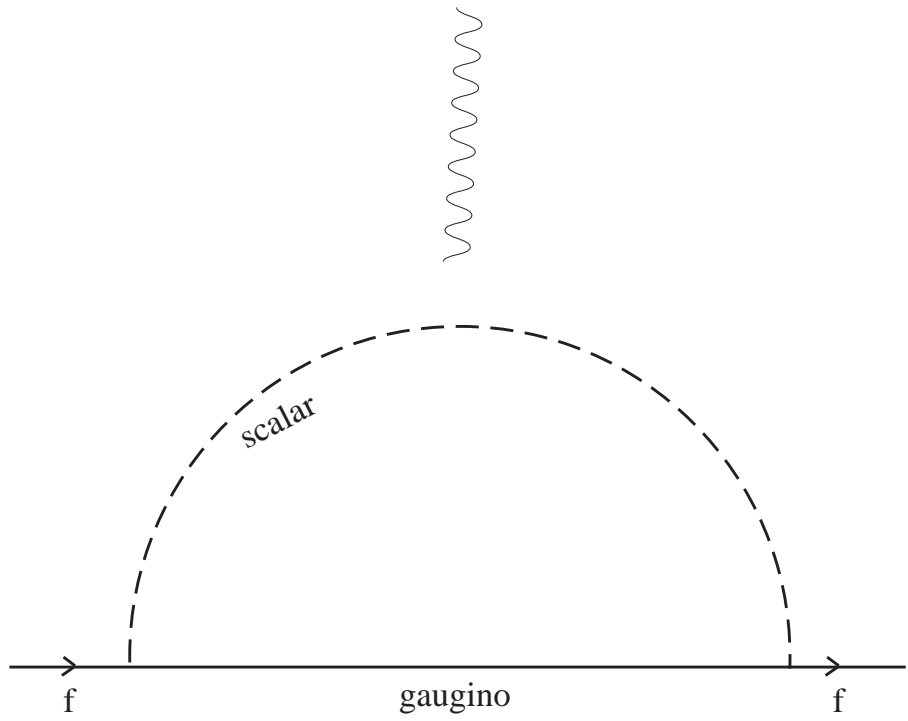

\centerline{ \DESepsf(edmfig1.epsf width 12 cm) }
\smallskip
\caption {\label{fig1} One loop diagram. The photon line can be attached
to any charged particle.}
\end{figure}

The EDM, $d_f$ for fermion f, appears in the effective Lagrangian $L_f$ as
\begin{eqnarray}  L_f&=&-{i\over 2}d_f\bar f\sigma_{\mu\nu}\gamma^5 f F^{\mu\nu}
\label{lag}
\end{eqnarray} The basic diagrams giving rise to $d_f$ are shown in
Fig.\ref{fig1}, and involve neutralino ($\tilde\chi^0_i$, i=1,4), chargino
($\tilde\chi^{\pm}_i$, i=1,2) and gluino ($\tilde g$) loops with squarks ($\tilde
q$) and sleptons ($\tilde l$). 

For the neutron dipole moment $d_n$ one must also take into account the gluonic
operators
\begin{eqnarray}  L^G&=&-{1\over
3}d^Gf_{abc}G_{a\mu\alpha}G_{b\nu}^{\alpha}\tilde{G}_c^{\mu\nu}
\label{lg}
\end{eqnarray} and
\begin{eqnarray}  L^C&=&-{i\over 2}d^C\bar
q\sigma_{\mu\nu}\gamma^5T^aqG_a^{\mu\nu}
\label{lc}
\end{eqnarray} where $\tilde{G}^{\mu\nu}_c= {1\over
2}\epsilon^{\mu\nu\alpha\beta}G_{c\alpha\beta}$, $f_{abc}$ are the SU(3)
structure constants and $T^a={1\over 2}\lambda^a$, where $\lambda^a$ are the
SU(3) Gell-Mann matrices. In addition to the diagrams of Fig.\ref{fig1}
with $\gamma$ replaced by $g$, 
operators of Eqs.(\ref{lag}-\ref{lc}) receive contributions from the two loop
Barr-Zee diagrams of Fig.\ref{fig2}\cite{ckp}, and the two loop Weinberg type
diagram of Fig.\ref{fig3} \cite{dai}. For the neutron dipole moment one must use
the QCD factors
$\eta^{ED}$, $\eta^G$, $\eta^C$ to evolve the results at the electroweak scale
down to 1 GeV \cite{aln}.

The calculation of $d_n$ suffers from QCD uncertainties from several sources.
These include:
\par\noindent (1) How to relate the quark moments $d_u$, $d_d$ to the neutron
moment
$d_n$. We use here the non relativistic quark model relation. (For other
approaches see Ref.\cite{bartl}.)
\begin{eqnarray}  d_n&=&{1\over 3}(4d_d-d_u)
\label{qm}
\end{eqnarray} (2) How to relate color and gluonic contributions to the electric
dipole moment. We use here the ``naive dimensional analysis'' of Ref.\cite{mon}.
\par\noindent (3) Uncertainty in $m_s$, the strange quark mass, which affects the
determination of $m_u$ and $m_d$. While the quark mass ratios are well known
\cite{lr},
\begin{eqnarray}  {m_u\over m_d}&=&0.553\pm 0.043;\,\,\,\, {m_s\over
m_d}=18.9\pm 0.8
\label{qr}
\end{eqnarray} one has $m_s=(175\pm 25)$MeV from QCD sum rules and $m_s=(100\pm
20 \pm 10)$ MeV  (2 GeV) from the quenched lattice calculation\cite{ms1} (Using
unquenched  approximation one expects an even smaller value \cite{ms2}).

Thus the values of $d_n$ calculated below have a significant uncertainty. In
spite of this, it is useful to see the effect on the SUSY parameter space of
simultaneously imposing the experimental constraints on both
$d_n$ and $d_e$ and we will discuss below how the uncertainty in $m_s$ effects
these constraints.

To specify our phase conventions, we give the mass matrices for the particles
entering in Fig.\ref{fig1}. With a convenient choice of phases, the chargino and
the neutralino mass matrices are:
\begin{eqnarray} M_{\chi^{\pm}}&=&\left(\matrix{
 \tilde m_2                & \sqrt 2 M_W sin\beta  \cr
  \sqrt 2 M_W cos\beta            &-|\mu|e^{i\theta} }\right)\label{char}
\end{eqnarray}

\begin{eqnarray} M_{\chi^0}&=&\left(\matrix{
 \tilde m_1  &0            &a   &b\cr
  0          &\tilde m_2   &c   &d  \cr
  a          &c            &0   &|\mu|e^{i\theta}\cr
  b          &d            &|\mu|e^{i\theta}   &0  \cr}\right)\label{neut}
\end{eqnarray}
  where $a=-M_Z sin\theta_W cos\beta$, $b=M_Z sin\theta_W sin\beta$,
$c=-cot\theta_W a$, $d=-cot\theta_W b$,
$tan\beta=v_2/v_1$ ($v_{1,2}=\mid <H_{1,2}>\mid$) and $\theta_W$ is the weak
mixing angle. The phase $\theta$ is given by
\begin{eqnarray}
\theta=\epsilon_1+\epsilon_2+\theta_\mu 
\end{eqnarray} where at the electroweak scale,
$<H_{1,2}>=v_{1,2}e^{i\epsilon_{1,2}}$, and
$\mu=|\mu|e^{i\theta_\mu}$. The squark mass matrices may be written as
\begin{eqnarray} M_{\tilde q}^2&=&\left(\matrix{
 m^2_{q_L}               & e^{-i\alpha_q}m_q(|A_q|+|\mu| R_q
e^{i(\theta+\alpha_q)}) 
\cr
  e^{i\alpha_q}m_q(|A_q|+|\mu| R_q e^{-i(\theta+\alpha_q)})            
&m^2_{q_R} }\right)\label{sqrk}
\end{eqnarray} where $m_q$, $e_q$ are the quark mass and 
charge,\begin{eqnarray} m_{q_L}^2&=&m^2_Q+m_q^2+(1/2-e_q sin^2\theta_W)M_Z^2
cos2\beta\\ m_{q_R}^2&=&m^2_U+m_q^2+e_q sin^2\theta_W M_Z^2 cos2\beta\label{qlqr}
\end{eqnarray}
$A_q=|A_q|e^{i\alpha_q}$, $m_Q^2$ and $m_U^2$ are given in Ibanez et al.
\cite{iba} and $R_q=cot\beta(tan\beta)$ for u(d) quarks. Similar expressions
hold for slepton masses, with phases $\alpha_l$. Our sign conventions for $A_q$
and $\mu$ are those of Ref.\cite{bop}.

The condition for electroweak symmetry breaking is obtained by minimizing the
effective potential $V_{eff}$ with respect to $v_1$, $\epsilon_1$, $v_2$ and
$\epsilon_2$. The Higgs sector of $V_{eff}$ is 
\begin{eqnarray}
V_{eff}&=&m_{1}^2v_1^2+m_{2}^2v_2^2+2|B\mu|cos(\theta+\theta_B)v_1v_2+
{g^2_2\over 8}(v_1^2+v_2^2)^2+{{g^{\prime}}^2\over
8}(v_2^2-v_1^2)^2+V_1\label{veff}
\end{eqnarray} where $V_1$ is the one loop contribution, $m^2_i=\mu^2+m^2_{H_i}$
and 
$m_{H_{1,2}}^2$ are the $H_{1,2}$ running masses. For particles of spin j one has
\begin{eqnarray}V_1={1\over {64 \pi^2}}\sum_a C_a(-1)^{2 j_a} (2 j_a+1) m_a^4
(ln{m_a^2\over Q^2}-{3\over 2})\label{v1}
\end{eqnarray}  where $C_a$ is the color degree of freedom of the $a^{th}$
particle. In the following we choose the low energy scale $Q$ to be $m_t$ (175
GeV), and
include the full third generation states, $t$, $b$ and $\tau$ in $V_1$. This
allows an examination of the large tan$\beta$ domain.

We view the minimization equations of $V_{eff}$ as equations to determine the
Higgs VEVs i.e. $v_1,\, v_2,\, \epsilon_1,\, \epsilon_2$. Thus in the tree
approximation, the extrema equations $\partial V_{eff}/\partial
\epsilon_i=0$
yield $2|B\mu| sin(\theta+\theta_B)=0$, and hence the minimum of $V_{eff}$
requires
\begin{eqnarray}\theta=\pi-\theta_B\label{theta}
\end{eqnarray} (the choice of $\theta=-\theta_B$ leads to a maximum). The one
loop corrections depend only on the mass eigenvalues, and hence from
Eq.(\ref{sqrk}) only on the phase $\theta+\alpha_q$ and $\theta+\alpha_l$. Thus,
at the one loop level, one gets a correction to Eq.(\ref{theta}) of the
form\cite{demir,carm} 
\begin{eqnarray}\theta=\pi-\theta_B+ f_1(\pi-\theta_B+\alpha_q,\pi-\theta_B+\alpha_l
)\label{thetaf}
\end{eqnarray}  where $f_1$ is the one loop correction with $\theta$
approximated by its tree  value, Eq.(\ref{theta}). As we will see, this
correction can become significant for large tan$\beta$. The EDMs however depend
on both the mass eigenvalues and the rotation matrices. Hence they will depend
on $\theta+\alpha_q$ and $\alpha_q$ separately, with $\theta$ determined by the
electroweak symmetry breaking condition in Eq.(\ref{thetaf}).

The current experimental 90$\%$ C.L. upper bounds on $d_n$\cite{dn} and
$d_e$\cite{de} are
\begin{eqnarray}(d_n)_{exp}&<&6.3\times 10^{-26}ecm;\,\,\,\,(d_e)_{exp}<4.3\times
10^{-27}ecm\label{dnde}
\end{eqnarray} In discussing the data, it is convenient to define the quantity
\begin{eqnarray}  K=log_{10}\mid{d_f\over {(d_f)_{exp}}}\mid
\label{k}
\end{eqnarray} Thus $K=0$ corresponds to a theoretical value which saturates the
current experimental bound, while $K=-1$ would represent the case, should the
experimental bounds be reduced by a factor of 10. As pointed out in
Refs.\cite{nath1,nath2,nath3,falk1,bk,falk2,kane1,kane2,bartl,pokorski},
various
cancellations can occur among the different contributions to the EDM. This is
illustrated in Fig.\ref{fig4} where $K$ is plotted vs. $m_0$ for the electron
dipole moment (eEDM).
\noindent We note that eventually for very large $m_0$, the curves fall below the
$K=0$ bound (as expected). Further the allowed range of $m_0$ (so that
$K\leq 0$) decreases with increasing tan$\beta$ (and the allowed range would
become very small should $K=-1$, i.e. the experimental bounds be reduced by a
factor of 10). In addition, the position of the dips moves to lower $m_0$ with
increasing tan$\beta$. This happens due to the fact that the regions on the right of
the dips are dominated by the contribution from the  chargino diagram. 
 As tan$\beta$ increases, the contribution from the  chargino diagram increases
much more than the contribution from the neutralino diagram. However, a decrease
in $m_0$ increases the neutralino  diagram much more than the chargino diagram.
As a result the dips shift towards the smaller $m_0$ value for larger 
tan$\beta$. We use negative values (arising from the factor $\pi$ in the Eq.
\ref{theta}) of $\mu$ to satisfy the experimental constraints
on the BR of $b\rightarrow s\gamma$ \cite{bsg1,bsg2}.

We begin our analysis by examining the RGE to determine what GUT scale parameters
lead to acceptable CP violating phases at the electroweak scale. We have used
Ref.\cite{bop} for the RGE relating the GUT scale parameters to the
parameters
at the electroweak scale. In general, the RGE must be solved numerically
and all
results given below are obtained by accurate numerical integration. However,
approximate analytic solutions can be found for low and intermediate tan$\beta$
(neglecting b and
$\tau$ Yukawa couplings) or for the SO(10) limit of very large tan$\beta$
(neglecting the $\tau$ Yukawa coupling). These analytic solutions give some
insight into the nature of the more general numerical results. 

The solution for the $A_t(t)$ parameter in the low and intermediate tan$\beta$
case  can be cast in the form \cite {landau}
\begin{eqnarray}A_0&=&{A_R(t)\over D_0}+{H_3(t)\over F(t)}m_{1/2}\label{a0}
\end{eqnarray} where $A_R$ is the residue at the t-quark Landau pole, 

\begin{eqnarray}A_R&=&A_t+m_{1/2}(H_2(t)-{H_3(t)\over {F(t)}})\label{ar}
\end{eqnarray}  and 
\begin{eqnarray}D_0&=&1-6{F(t)\over E(t)}Y(t)\label{d0}.
\end{eqnarray} Here $t=2ln(M_G/Q)$, the form factors $E$, $F$, $H_2$ and $H_3$
are real and are defined in
\cite{iba} and $Y(t)=h_t/{16 \pi^2}$, where $h_t$ is the t-quark Yukawa coupling
constant. $D_0$ vanishes at the t-quark Landau pole (for Q=$m_t$,
$D_0\cong 1-(m_t/{200 sin\beta})^2)$ and is generally small (i.e.
$D_0\stackrel{<}{\sim} 0.2$ for $m_t=175$ GeV). The imaginary part of
Eq.(\ref{a0}) gives:
\begin{eqnarray}D_0|A_0|sin\alpha_{0A}&=&|A_t|sin\alpha_t\label{d0a0}.
\end{eqnarray} Thus even if $\alpha_{0A}=\pi/2$, $\alpha_t$ at the electroweak
scale will be generally suppressed due to the smallness of $D_0$, i.e. the RGE
naturally make the phase $\alpha_t$ small due to the nearness of the Landau
pole.
(This result has been previously observed, for low tan$\beta$ in 
\cite{falk1}.) The approximate SO(10) solution with large tan$\beta$
(where 
$Y_t\cong Y_b$) gives a similar analytic form for $A_t(t)\cong A_b(t)$  with the
factor 6 replaced by 7 in Eq.(\ref{d0}). Thus the suppression effect on
$\alpha_t$ occurs over the entire tan$\beta$ domain. These effects thus allow
$\alpha_{0A}$ to be large without a priori violating the experimental EDM bounds.

In contrast there is no analogous RGE suppression effect in the first generation
$A$ parameters $A_u$, $A_d$ and $A_e$. From the RGE, one finds that
$A_u$ has the form of Eq.(\ref{a0}) for low and intermediate tan$\beta$ with the
factor 6 replaced by 1 in Eq.(\ref{d0}). Now $D_0\simeq 1$ and no suppression
effect occurs for $\alpha_u$ at the electroweak scale in the analogue of
Eq.(\ref{d0a0}). However, the EDMs are not very sensitive to the first
generation $\alpha_u$, $\alpha_d$, $\alpha_e$ and one can have large values of
these parameters without violating the experimental bounds in a reasonable
region of the SUSY parameter space.

From the RGE, one sees that the phase of the $\mu$ parameter does not run at the
1-loop level, i.e.
$\theta_{0\mu}=\theta_{\mu}$, where $\theta_{\mu}$ is the phase at the
electroweak scale. Thus a large GUT scale phase will lead to a large electroweak
scale phase. However, as seen in Eqs.(\ref{char}-\ref{sqrk}),
$\theta_\mu$ enters only in the combination $\theta$, and $\theta$ is determined
in terms of $\theta_B$, $\alpha_l$ and $\alpha_q$ by the minimization of the
effective potential, as given in Eq.(\ref{thetaf}). Thus (with our choice of the
phases in the mass matrices) $\theta_{\mu}$ does not enter in the EDMs.

The EDMs, however are highly sensitive to $\theta_B$, which by
Eq.(\ref{thetaf}), enters into all the diagrams, and over most of the parameter
space, $\theta_B$ must be small to satisfy the EDM experimental bounds. This is
illustrated in Fig.\ref{fig5} where we have plotted the eEDM $K$ vs.
$\theta_B$. We see that $\theta_B$ can be large i.e. $\theta_B\simeq 0.08$, for
low $\tan\beta=3$, though it becomes smaller for higher
$\tan\beta$. However, even for tan$\beta=3$, the range $\Delta\theta_B$ of
$\theta_B$ so that $K\leq 0$ is very small i.e.
$\Delta\theta_B\stackrel{<}{\sim} 0.02$ (and $\theta_B$ would become quite fine
tuned if the experimental bound were reduced by a factor of 3 i.e.
$K=-0.5$). Figs.6 show that $\Delta\theta_B$ remains small for
$m_{1/2}\stackrel{<}{\sim} 350$ GeV ($m_{\tilde g}\stackrel{<}{\sim} 1$ TeV) and
only relaxes somewhat when $m_{1/2}\simeq 700$ GeV ($m_{\tilde
g}\stackrel{<}{\sim}2$ TeV). In this figure we have plotted the $K\le 0$
regions
as a function of $\theta_B$ and
$m_{1/2}$ for three different values of tan$\beta$. Since the contribution from
the chargino diagram increases with tan$\beta$, the larger tan$\beta$ requires
a smaller phase to produce the necessary cancellation between the chargino
and the
neutralino diagrams.

The RGE allow us to examine the significance of the above results at the GUT
scale. The RGE for $B$ and $A_t$ can be solved in the low and intermediate
tan$\beta$ regime to obtain
\begin{eqnarray}B&=&B_0-{1\over 2}(1-D_0)A_0-\Phi m_{1/2}\label{b}
\end{eqnarray}  where 
\begin{eqnarray}\Phi&=&-{1\over 2}(1-D_0){H_3\over F}+ [3h_2+{3\over
5}h_1]{\alpha_G\over{4\pi}}
\label{phi}
\end{eqnarray} 
$h_i=t/(1+\beta_it)$, $\alpha_G\cong 1/24$ is the GUT gauge coupling  constant
and $\beta_i$ are the MSSM beta functions. The real and the imaginary parts of
Eq.(\ref{b}) give: \begin{eqnarray}|B|sin\theta_B&=&|B_0|sin\theta_{0B}-{1\over
2}(1-D_0)|A_0|sin\alpha_{0A}\label{bsin}
\end{eqnarray}
\begin{eqnarray}|B|cos\theta_B&=&|B_0|cos\theta_{0B}-{1\over
2}(1-D_0)|A_0|cos\alpha_{0A}-\Phi m_{1/2}\label{bcos}
\end{eqnarray} Eqs.(\ref{bsin},\ref{bcos}) can be viewed as determining the
electroweak scale values of $|B|$ and $\theta_B$ in terms of the GUT input
parameters. Thus
$\theta_B$ depends upon both the initial phase $\theta_{0B}$ and the $A_0$ phase
$\alpha_{0A}$. Alternatively, one may use Eqs.(\ref{bsin},\ref{bcos}) to impose
electroweak scale phenomenological constraints on the allowed GUT parameters. 
The requirement that the GUT theory gives rise to electroweak symmetry breaking
gives 
\begin{eqnarray}|B|&=&{1\over 2}sin{2\beta}{m_3^2\over|\mu|}\label{bmag}
\end{eqnarray} where $m_3^2=\mu^2_1+\mu^2_2$ and
$\mu_i^2=|\mu|^2+m_{H_i}^2+\Sigma_i$. (Here
$m_{H_i}^2$ are the running Higgs masses at the electroweak scale, $|\mu|$ is
determined by electroweak breaking and $\Sigma _i$ are the one loop corrections
\cite{ap}.) In addition, we have seen that the experimental  bounds on the EDMs
restrict $\theta_B$ to be small, i.e.
$|\theta_B|\stackrel{<}{\sim} 0.1$ (and usually much smaller) and the allowed
range,
$\Delta\theta_B$, that satisfies the EDMs is quite small. These conditions put
severe constraints on the GUT scale theory which we now discuss. 

Consider first the case where tan$\beta$ is very close to its minimum value
e.g. tan$\beta$=2. Here $1/2 sin2\beta$ is not small and by   Eq.(\ref{bmag}),
$|B|$ is of normal size. Eq.(\ref{bsin}) then implies that $\theta_B$ and
$\theta_{0B}$ are of roughly the same size (even for $\alpha_{0A}=\pi/2$) as
previously noted in Ref.\cite{falk1}. Thus a reasonable GUT theory
requires one
only to justify the size of 
$\theta_{0B}$. However, as tan$\beta$ increases, $1/2 sin2\beta$
 decreases, and unless $\mu$ becomes anomalously small or $m_3^2$ anomalously
large, Eq.(\ref{bmag}) implies that $|B|$ becomes small. Thus the lefthand side
of Eq.(\ref{bsin}) becomes small (and to first approximation can be neglected),
and Eq.(\ref{bsin}) implies that $\theta_{0B}$ is determined by $\alpha_{0A}$
mainly, i.e. $\theta_{0B}$ is large if
$\alpha_{0A}$ is large. This result is illustrated in Fig.\ref{fig7} where
$\theta_{0B}$ is plotted vs. $\theta_B$. We see that since $\theta_B$ is small
for all tan$\beta$, $\theta_{0B}$ is in general large, e.g. even for tan$\beta
=3$ one has $\theta_{0B}\simeq 0.8$.

Returning to Eq.(\ref{bsin}), we see, however, that $\Delta\theta_{0B}$, the
range of values of $\theta_{0B}$ that will satisfy the EDM constraints, is very
small. Thus for fixed $A_0$, Eq.(\ref{bsin}) implies for large tan$\beta$ that
\begin{equation}
\Delta\theta_{0B}\cong{|B|\over {|B_0|}}\Delta\theta_B\ll\Delta\theta_B
\label{DthetaB}
\end{equation} 
and since $\Delta\theta_B$ is small, $\Delta\theta_{0B}$ will be
very small. This is illustrated in Fig.\ref{fig8} where the allowed values of
$\theta_{0B}$ for $K\leq 0$ are plotted vs. $m_{1/2}$ for $d_e$ and $d_n$ for
tan$\beta$=3, 10, 20. We see that the phenomenological constraints at the
electroweak scale imply that $\theta_{0B}$ is both large and its value is
sharply fine tuned (unless $\alpha_{0A}$ is small, SUSY masses are large or
tan$\beta$ is small). One may alternatively view this from the ``top down'': if
Planck physics determines $\alpha_{0A}$ and $\theta_{0B}$ to be large and fixed,
then if the model is to achieve electroweak symmetry breaking with EDMs below
existing bounds, there will be a fine tuning of other GUT scale parameters
(unless tan$\beta$ is small or the SUSY masses are large).

As discussed above, there are a number of uncertainties in the calculation of
the neutron EDM. Fig.\ref{fig9} illustrates the effects of varying the d-quark
mass in a plot of the region allowed by the experimental EDM bounds ($K\leq 0$)
in the $m_0-m_{1/2}$ plane for $\theta_B=0$, $|A_0|=300$ GeV,
$\alpha_{0A}=\pi/2$, tan$\beta$=3. The bound on $d_e$ already excludes all
regions below the lower solid curve, while the $d_n$ bound excludes regions to
the left of the upward running curves for $m_d=5$ MeV (dotted), $m_d=8$ MeV
(solid), $m_d=12$ MeV (dashed). Thus the combined exclusion region increases from
$m_{1/2}\stackrel{>}{\sim} 260$ GeV to $m_{1/2}\stackrel{>}{\sim} 440$ GeV as
$m_d$ increases, a significant change. In the other figures in this paper we
have used the middle value of $m_d=8$ MeV corresponding to $m_s\simeq 150$ MeV. 

In Fig.\ref{fig9}, we have chosen $\theta_B=0$. As one increases
$\theta_B$, one will eventually arrive at a neutralino-chargino cancellation
region, and this will reduce the values of $m_0$ and $m_{1/2}$ which are
excluded. Further, for low tan$\beta$ this cancellation can occur at relatively
large $\theta_B$ (i.e. $\theta_B\simeq 0.2$) as shown in Ref.\cite{falk1}.
Alternatively, increasing tan$\beta$ can also cause this cancellation, as the
loop corrections, Eq.(\ref{thetaf}), become large with large tan$\beta$ and the
contribution from the  chargino diagram increases with tan$\beta$.  This is
illustrated for tan$\beta$=20 in Fig.\ref{fig10} where the $d_n$ curve for $K=0$
(solid) bends downward below $m_0$=500 GeV when 200 GeV$<m_{1/2}<400$ GeV,
showing in this exceptional type situation that one can satisfy the EDM constraints
with a light particle spectrum and large tan$\beta$. Note that the cancellation
can occur for a relatively wide band of $m_{1/2}$, and persists even for
$K=-0.5$.

The regions of the SUSY parameter space that get eliminated by a joint
imposition of the experimental bounds on the $d_e$ and $d_n$ are sensitive, of
course, to the choice of the parameters. Thus in Figs.\ref{fig6}  one sees that
for $m_0=100$ GeV, one requires $m_{1/2}\stackrel{>}{\sim} 350$ GeV ($m_{\tilde
g}\stackrel{>}{\sim} 1$TeV) to jointly satisfy the $d_e$ and $d_n$ bounds, and
this requirement is roughly independent on tan$\beta$. Fig.\ref{fig11} shows
that if $m_0$ is increased to 250 GeV one requires now
$m_{1/2}\stackrel{>}{\sim} 160$ GeV ($m_{\tilde g}\stackrel{>}{\sim} 450$ GeV)
to satisfy both bounds, i.e. if $m_0$ is increased, the lower bound on $m_{1/2}$
is decreased as expected. However, if $|A_0|$ is increased, it is not necessary
that the allowed domain of $m_{1/2}$ increases. This is illustrated in
Fig.\ref{fig12} for the choice $|A_0|$=800 GeV, $m_0$=100 GeV. The increase of
$|A_0|$ raises the allowed values of $\theta_B$ for both $d_e$ and $d_n$,
but the raising of $d_e$ is larger causing the joint
allowed domain to require $m_{1/2}\stackrel{>}{\sim}  550$ GeV
($m_{\tilde{g}}\stackrel{>}{\sim} 1.5$ TeV). 

In order to exhibit the size of $\mid\mu\mid$, we have plotted
$\mid\mu\mid$ as a function of $m_{1/2}$ for $m_0$=100, 300 and 700 GeV in
Fig.\ref{fig13}. The EDM constraints have not been imposed in
Fig.\ref{fig13}, but from Fig.6a, one sees that they are
satisfied at least for $m_0$=100 GeV for $d_e$ for essentially the entire
$m_{1/2}$ range (which allows $\theta_B\simeq 0.08$). Note that
$\mid\mu\mid$ is large (i.e. $\mid\mu\mid^2/M_Z^2\gg 1$) over the full
$m_{1/2}$ domain so that one is generally in the gaugino scaling domain.

\section{Nonuniversal soft breaking}

There are three types of nonuniversalities that might be considered in SUGRA
models: nonuniversal gaugino masses at $M_G$, nonuniversal scalar masses at
$M_G$ and generation off diagonal cubic soft breaking matrices
$A_{ij}$ and scalar mass matrices $m^2_{ij}$. We will discuss here briefly the
first two possibilities, a more detailed analysis will be given in a later paper
\cite{aad}. (The third type of non universality has the possibility of
generating $\epsilon$ and $\epsilon^{\prime}/\epsilon$ CP violations independent
of the CKM phase \cite{epsilon}.)

In general, the U(1), SU(2) and SU(3) gaugino masses at $M_G$ can have the form
\begin{eqnarray} m_{1/2i}&=&|m_{1/2i}|e^{i\phi_i};\, i=1,2,3
\label{mg}
\end{eqnarray} A convenient phase choice is to set the SU(2) phase to zero:
$\phi_2=0$, and hence the chargino mass matrix, Eq.(\ref{char}), is
unchanged. The remaining phases produce effects in the RGE for $A_t(t)$:
\begin{eqnarray} A_t&=&D_0A_{0}+\sum_i\Phi_{ti}|m_{1/2i}|e^{i\phi_i}
\label{At}
\end{eqnarray} and hence the imaginary part gives
\begin{eqnarray}
|A_t|sin\alpha_t&=&|A_0|D_0sin\alpha_{0A}+\sum_i\Phi_{ti}|m_{1/2i}|sin\phi_i
\label{At2}
\end{eqnarray} Thus the Landau pole factor $D_0$ still suppresses any large phase
$\alpha_{0A}$ at the GUT scale, though $\alpha_t$ may become large due to the
$\phi_i$ phase which does not run at one loop order and hence is not suppressed.
This phenomena can affect $d_n$.

A second effect occurs in the neutralino mass matrix which now reads
\begin{eqnarray} M_{\chi^0}&=&\left(\matrix{
 |\tilde m_1|e^{i\phi_1}  &0            &a   &b\cr
  0          &\tilde m_2   &c   &d  \cr
  a          &c            &0   &|\mu|e^{i\theta}\cr
  b          &d            &|\mu|e^{i\theta}   &0  \cr}\right)\label{neut2}
\end{eqnarray} where $\theta =\epsilon_1+\epsilon_2+\theta_\mu$. Thus $\phi_1$
effects any neutralino-chargino cancellations in $d_e$ and $d_n$. An example is
illustrated in Fig.\ref{fig14}. One sees that $\theta_B$ can be large but is
still tightly constrained to satisfy the experimental EDM bound (i.e.
$\Delta\theta_B\simeq 0.02$). One is again led to a fine tuning at the GUT
scale.

As an example of nonuniversal sfermion masses, we consider an SU(5)-type model
where the GUT group possesses an SU(5) subgroup with matter embedded in the
usual way in $10+\bar 5$ representations. We also assume the first two
generations remain universal to suppress flavor changing neutral currents. The
Higgs masses have the form at $M_G$
\begin{eqnarray}
m_{H_1}^2&=&m_0^2(1+\delta_1);\,\,\,\,m_{H_2}^2=m_0^2(1+\delta_2)
\label{scanonuni}
\end{eqnarray} while the third generation $10$ representation (containing
$\tilde t_L$,
$\tilde t_R$,$\tilde b_L$, $\tilde\tau_R$) and the $\bar 5$ (containing
$\tilde b_R$,
$\tilde\nu_\tau$, $\tilde\tau_L$) masses are parametrized by
\begin{eqnarray} m_{10}^2&=&m_0^2(1+\delta_{10});\,\,\,\,m_{\bar
5}^2=m_0^2(1+\delta_5)
\label{scanonuni1}
\end{eqnarray} Here the $\delta_i\ge-1$ represent the deviations from the
universality. In addition the third generation cubic soft breaking parameters are
$A_{0t}$ and $A_{0b}=A_{0\tau}$.

One may use the non universal sfermion and Higgs masses to soften the effects of
the experimental EDMs. This is illustrated in Fig.\ref{fig15} where the K=0
constraint is imposed in the $m_0-m_{1/2}$ plane for a choice of
$\delta_i$ parameters. For $\delta_i=0$, the $d_n$ curve would continue to rise
for small $m_{1/2}$ as seen in Fig.\ref{fig9}, rather than turn over as in
Fig.\ref{fig15}. For example, for $m_{1/2}$=220 GeV and tan$\beta$=5, 
$K(d_n)<0$ occurs for $m_0>580$ GeV
in the non universal case, 
whereas $K(d_n)<0$ occurs for $m_0>750$ GeV in the universal case. 
The particular set of 
$\delta_i$s in the Fig.\ref{fig15} is chosen to reduce the stop and the sbottom
mass and to 
satisfy the requirement of the radiative electroweak symmetry breaking.
 The lower stop and sbottom mass increases the Weinberg type diagram which has a subtractive
 effect in the net nEDM magnitude.

\section{SUGRA Model Of Small Phases}

While choosing the CP violating phases to be small, i.e. $\phi_i\simeq 10^{-2}$,
to satisfy the EDM constraints appears to be artificial in the low energy MSSM,
it is possible that a natural model of this type can arise in SUGRA GUT models. 

SUGRA models depend on three functions of the chiral fields
${\Phi_{\alpha}}$: $f_{\alpha\beta}(\Phi_{\alpha} )$ the gauge kinetic energy,
$K(\Phi_{\alpha} ,\Phi_{\alpha}^{\dag} )$ the Kahler potential and
$W(\Phi_{\alpha} )$ the superpotential. One assumes a hidden sector exists where
some fields, e.g. moduli or dilaton, $\Phi_i$ grow VEVs of Planck size to break
supersymmetry i.e.
\begin{eqnarray} x_i&=&\kappa <\Phi_i>=O(1)
\label{xi}
\end{eqnarray} where $\kappa^{-1}=M_{Pl}=2.4\cdot 10^{18}$ GeV. Thus one can
write
$\{\Phi_{\alpha}\}=\{\Phi_i, \Phi_a\}$ where $\Phi_a$ are the physical
sector
fields. One might expand the Kahler potential in a power series of the physical
fields:
\begin{eqnarray} K&=&\kappa^{-2}c^{(0)}+(c_{ab}^{(2)}\Phi_a\Phi_b+{1\over M}
c_{abc}^{(3)}\Phi_a\Phi_b\Phi_c +...) +(\tilde c_{ab}^{(2)}\Phi_a\Phi_b^{\dag}
+{1\over M}                  
\tilde c_{abc}^{(3)}\Phi_a^{\dag}\Phi_b\Phi_c+...)
\label{kahler}
\end{eqnarray} where $M$ is a large mass. The $c^{(i)}, \tilde c^{(i)}$ are
dimensionless and we assume them to be $O(1)$. The first parenthesis is
holomorphic, and can be transferred to the superpotential by a Kahler
transformation giving rise to the $\mu$ term:
\begin{equation} W\rightarrow W +
\kappa^2W(c_{ab}^{(2)}\Phi_a\Phi_b+{1\over                  
M}c_{abc}^{(3)}\Phi_a\Phi_b\Phi_c+...)
\label{W}
\end{equation} The leading additional terms on the right occur when $W$ there is
replaced by its
VEV ($\kappa^2<W>\simeq m_{3/2}$) and one of the fields in the cubic term,
e.g.
$\Phi_c$, has a GUT scale VEV. Then $W\rightarrow W+\mu_{ab}\Phi_a\Phi_b$,
where 
\begin{eqnarray}
\mu_{ab}&=&(c_{ab}^{(2)}+{M_G\over M}c_{abc}^{(3)})m_{3/2}
\label{muab}
\end{eqnarray} In the perturbative heterotic string we expect $M\simeq M_{Pl}$
and hence
$M_G/M\simeq O(10^{-2})$. If one assumes $c^{(2)}$ is real and $c^{(3)}$ is
complex with arbitrary size phase, then $\mu$ naturally has a phase of size
$\theta_{0\mu}\simeq O(10^{-2})$. A similar analysis can be done for the other
parameters (A, B, $\tilde m$) yielding automatically small phases. One assumes
only that the leading (renormalizable) term in the matter expansion is real,
while the higher terms (presumably arising from integrating out the tower of
massive states) can have arbitrary phases.

Other string scenarios can have $M\simeq O(M_G)$. In that case one expects the
nonrenormalizable terms to give rise to large phases, unless the low lying
members of the tower of heavy states do not couple to the fields of the physical
sector.

\section{Conclusions}

In SUGRA GUT models, the basic parameters are given at the GUT scale. The
renormalization group equations then carry this information to the electroweak
scale, allowing for experimental tests of the model. Thus what represents a
``natural'' choice of parameters is presumably a GUT scale question.

The experimental values of the electric dipole moments put strong constraints on
the low energy SUSY parameter space. However, for mSUGRA models at least, the
RGE suppress the value of the $A_t$ phase at the electroweak scale, $\alpha_t$,
so that any size of phase, $\alpha_{0A}$, even $\pi/2$ at
$M_G$ will generally lead to acceptable EDMs. This effect is due to the nearness
of the t-quark to its Landau pole.

The EDMs, however, are very sensitive to the B phase at the electroweak scale,
$\theta_B$. In the domain of light SUSY mass spectrum (i.e.
$\stackrel{<}{\sim}$1 TeV) and $\alpha_{0A}\simeq O(1)$, the EDMs allow a
large
$\theta_B$ (i.e.
$O(10^{-1}))$ only for low tan$\beta$, i.e. tan$\beta\stackrel{<}{\sim}3$.
Further, for fixed SUSY parameters in this domain, the allowed range
$\Delta\theta_B$ of
$\theta_B$ needed to satisfy the EDM constraints is very small for
tan$\beta\stackrel{>}{\sim}3$.

The above results strongly affect the GUT theory. Thus for $\alpha_{0A}$ large
and a light mass spectrum one finds at the GUT scale that $\theta_{0B}$ is
generally large, i.e. $O(1)$ even for low tan$\beta$, and the allowed range
$\Delta\theta_{0B}$ for satisfying the EDMs is exceedingly small for
tan$\beta\stackrel{>}{\sim} 3$. Thus for this situation, $\theta_{0B}$ must be
chosen large and be very precisely determined as tan$\beta$ gets large, leading
to a new fine tuning problem at the GUT scale. The origin of this difficulty
resides in the requirement that the GUT models give rise to parameters at the
electroweak scale that simultaneously satisfy the experimental EDM constraints
and give rise to radiative electroweak breaking.

The above discussion suggests that reasonable mSUGRA models with large phases
and mass spectrum below 1 TeV can only be constructed for small tan$\beta$ i.e.
tan$\beta\stackrel{<}{\sim}3$. The current LEP data combined with
cosmological constraints already require
tan$\beta\stackrel{>}{\sim} 2$ \cite{lep}, and LEP, the Tevatron and the LHC
 will be
able to explore the domain tan$\beta\stackrel{<}{\sim} 50$\cite{wag}. Thus the value of tan$\beta$ (or a
lower bound) is something that will be determined experimentally in the not too
distant future. Should tan$\beta$ turn out to be large, it is possible that only
models with small GUT phases $\phi_{0i}\simeq O(10^{-2})$ are reasonable, and it
was shown above that such models can arise naturally with $\phi_{0i}\simeq
O(M_G/M_{Pl})$.

Our analysis has included the loop corrections to the radiative breaking
condition. While these corrections are small, they grow with tan$\beta$
and are competitive to the $\theta_B$ contribution to the EDMs. Thus for
large tan$\beta$ (e.g. tan$\beta\simeq 20$) they can produce a new
cancellation phenomena (with $\theta_B$=0), allowing the EDM constraints to be satisfied for a
light mass spectra (e.g. $m_0\simeq$300 GeV, $m_{1/2}\simeq$ 200 GeV)
which for lower tan$\beta$ would be excluded by the EDM data.

The combined restrictions of the experimental EDM bounds for the neutron
and electron put strong constraints on the SUSY parameter space. However,
since the theoretical cancellations needed to satisfy experiment are
delicate, as one varies the parameters one can get rather different
excluded regions. Thus, while we have found that increasing $m_0$ gives an
allowed region of smaller $m_{1/2}$ (as one might have expected),
increasing $\mid A_0\mid$ can require $m_{1/2}$ to be increased to satisfy
the EDM bounds. However, it should be stressed that there are significant
theoretical uncertainties in the calculation of $d_n$, and we have seen,
for example, that the current uncertainty in the quark masses $m_u$ and
$m_d$ (arising from the uncertainty in $m_s$) can lead to a significant
variation in the domain the neutron EDM data can exclude.

Much of the above discussion holds also for non minimal models with non
universal gaugino and sfermion soft breaking masses. A more detailed discussion
of the effects of non universalities will be given elsewhere
\cite{aad}.

\section{Acknowledgement}

This work was supported in part by National Science Foundation Grant No.
PHY-9722090. One of us (R.A.) is pleased to acknowledge the hospitality of the
Korea Institute for Advanced Study where part of this paper was written.
 We would also like to thank Toby Falk and Keith Olive for valuable discussions.

\begin{figure}[htb]
\centerline{ \DESepsf(edmfig2.epsf width 12 cm) }
\smallskip
\caption {\label{fig2} Two loop Barr-Zee diagrams. A is the CP odd Higgs and
$\tilde q_i$ are the mass diagonal squark states.}
\end{figure}\begin{figure}[htb]
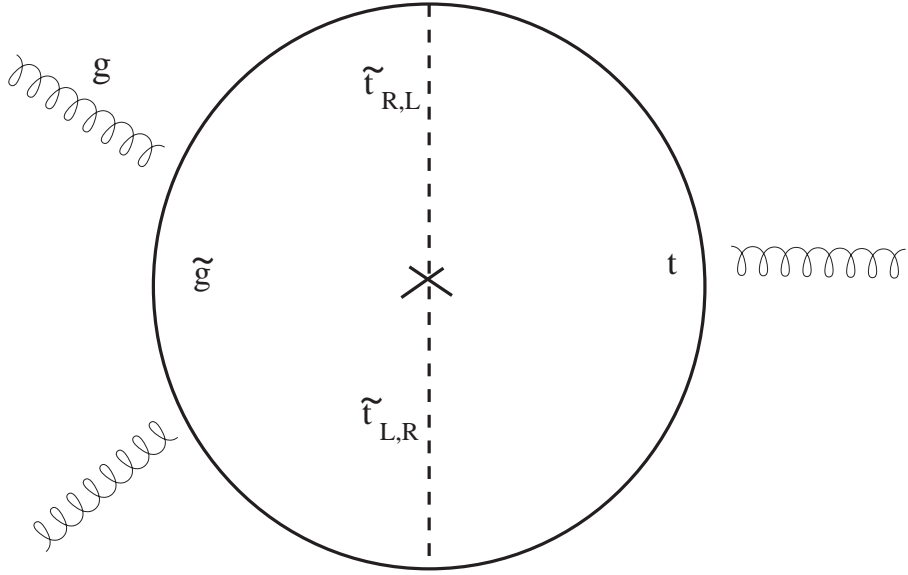

\centerline{ \DESepsf(edmfig3.epsf width 12 cm) }
\smallskip
\caption {\label{fig3} Two loop Weinberg type diagram.}
\end{figure}
\begin{figure}[htb]
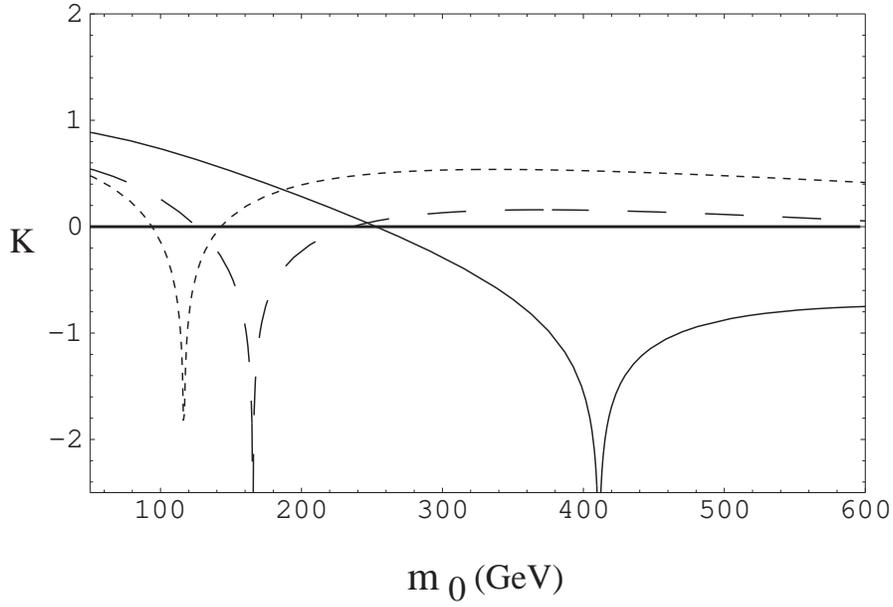
 
\centerline{ \DESepsf(edmfig4.epsf width 12 cm) }
\smallskip
\caption {\label{fig4} K (defined in the text) is plotted as a function of $m_0$
for $d_e$. The solid, dashed and dotted lines are  for tan$\beta=3$, 10 and
20 respectively. The other input  parameters are
$\alpha_{0A}={\pi\over 2}$, $|A_0|= 300$ GeV, $\theta_B=$0.02 and $m_{1/2}=300$
GeV. }
\vspace{0 cm}
\end{figure}\begin{figure}[htb] 
\centerline{ \DESepsf(edmfig5.epsf width 12 cm) }
\smallskip
\caption {\label{fig5} K (for eEDM) is plotted as a function of $\theta_B$. The
solid, dashed and dotted lines are  for tan$\beta=3$, 10 and 20 respectively.
The other input parameters are
$\alpha_{0A}={\pi\over 2}$, $|A_0|= 300$ GeV, $m_{1/2}=$ 300 GeV and
$m_0$= 100 GeV.}
\end{figure}
\begin{figure}[htb] 
\centerline{ \DESepsf(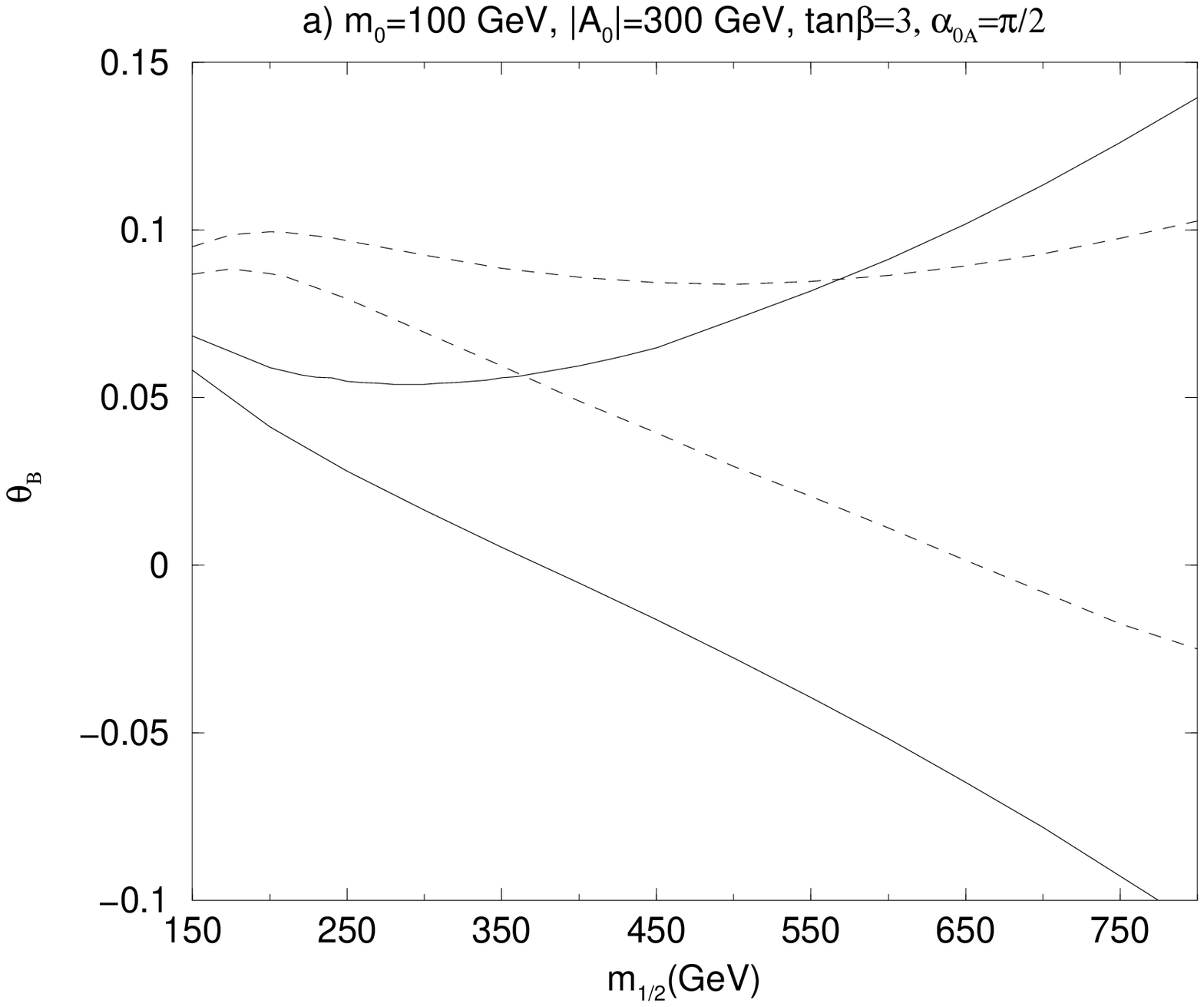 width 12 cm) }
\centerline{ \DESepsf(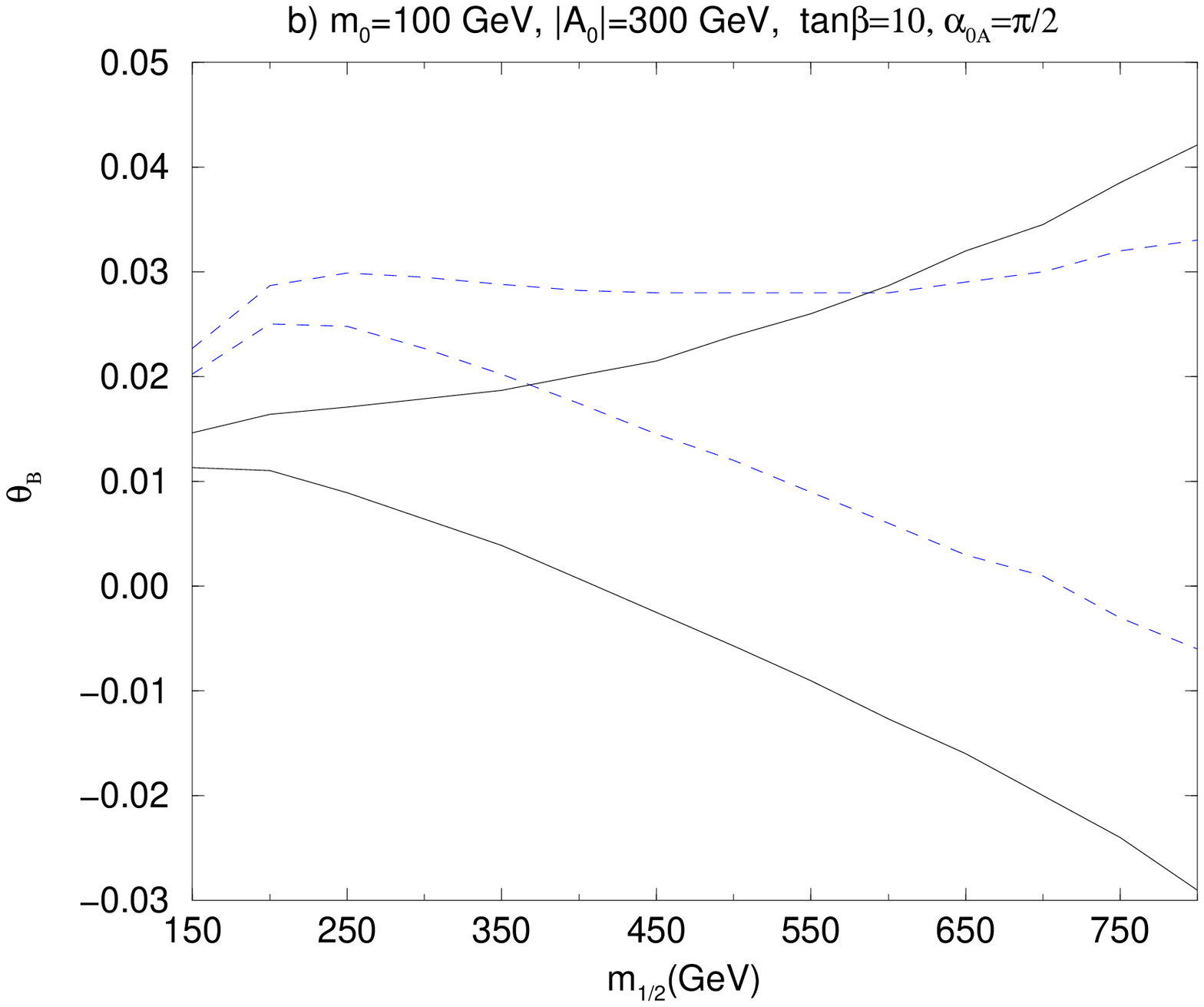 width 12 cm) }
\centerline{ \DESepsf(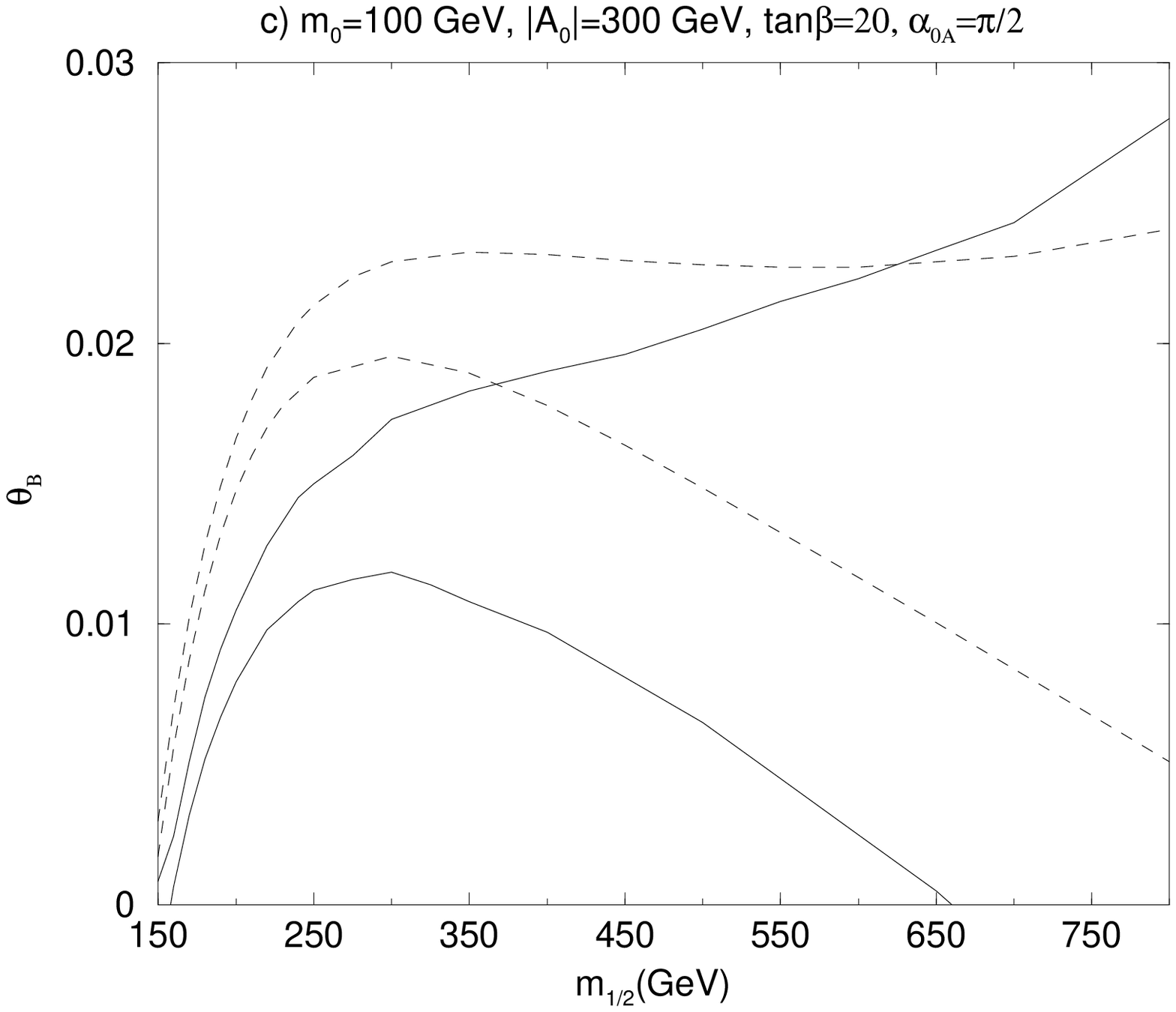 width 12 cm) }
\smallskip
\caption {\label{fig6} The K=0 contours plotted as a function of 
$\theta_B$ and $m_{1/2}$. The solid lines are for  $d_n$ and the dotted lines
are for $d_e$.}
\vspace{0 cm}
\end{figure}\begin{figure}[htb] 
\centerline{ \DESepsf(edmfigthetab1.epsf width 12 cm) }
\centerline{ \DESepsf(edmfigthetab2.epsf width 12 cm) }
\smallskip\smallskip
\caption {\label{fig7}$\theta_{0B}$ is plotted as a function of $\theta_B$. 
 Figs. a, b and c are for tan$\beta=3$, 10 and 20 respectively. The other input 
parameters are
$\alpha_{0A}={\pi\over 2}$, $|A_0|= 300$ GeV, $m_{0}=100$ GeV and 
$m_{1/2}=300$ GeV.}
\vspace{0 cm}
\end{figure}\newpage
\begin{figure}[htb] 
\centerline{ \DESepsf(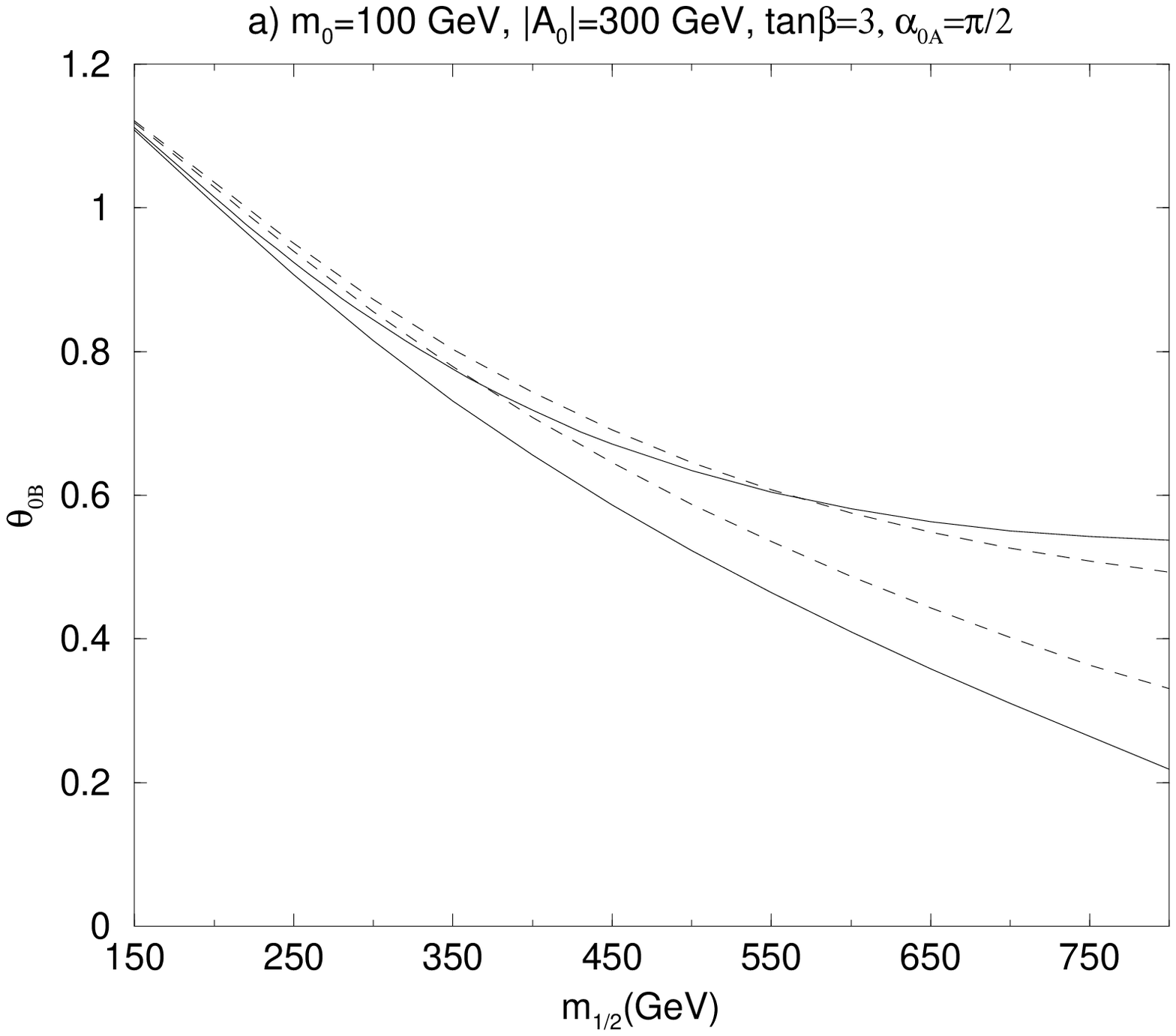 width 12 cm) }
\centerline{ \DESepsf(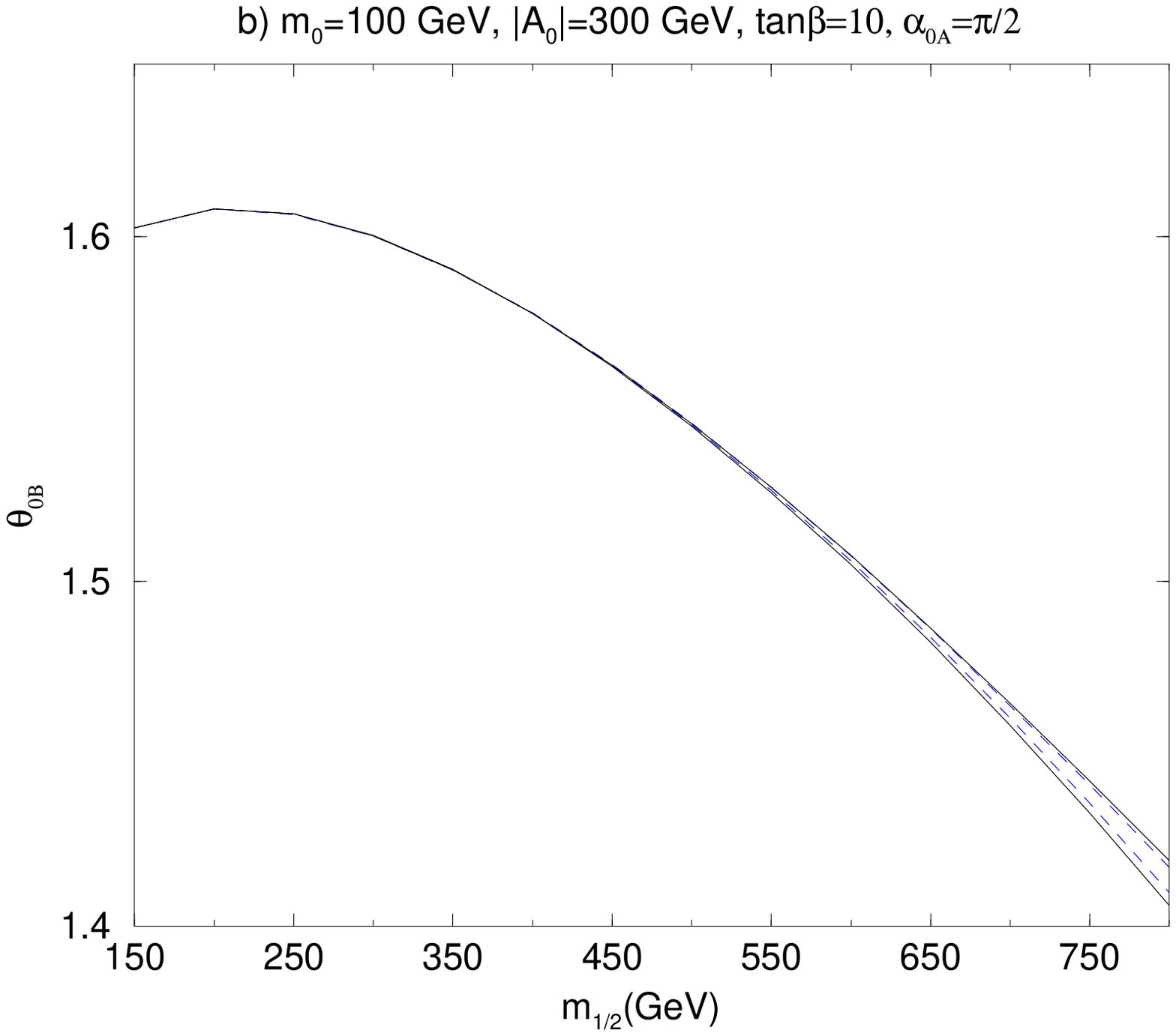 width 12 cm) }
\centerline{ \DESepsf(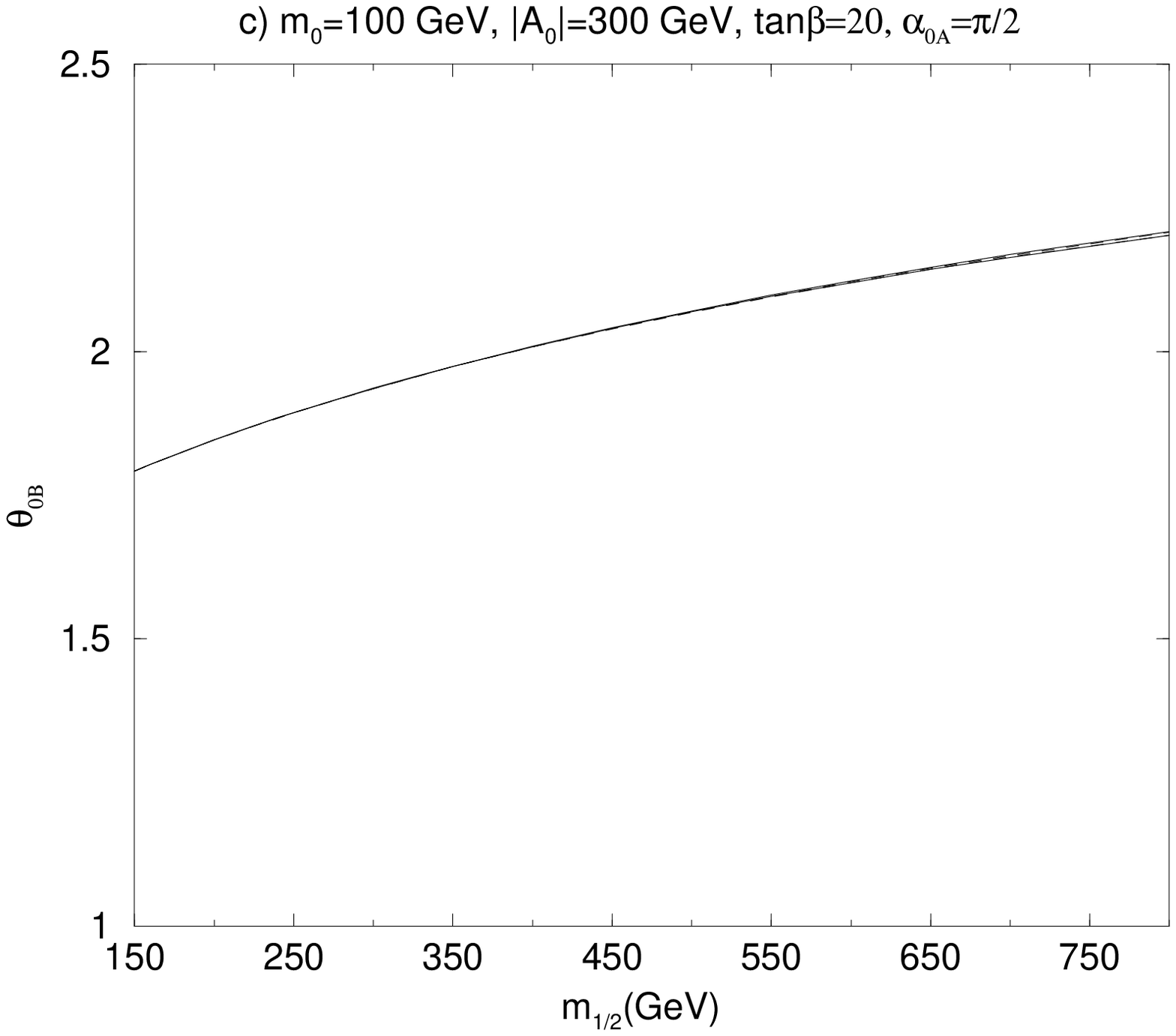 width 12 cm) }
\smallskip
\caption {\label{fig8}$\theta_{0B}$ vs $m_{1/2}$. Upper and lower lines are the
allowed range so that $K\le 0$. The solid lines are for  $d_n$ and the dotted
lines are for $d_e$.}
\vspace{0 cm}
\end{figure}
 \begin{figure}[htb]
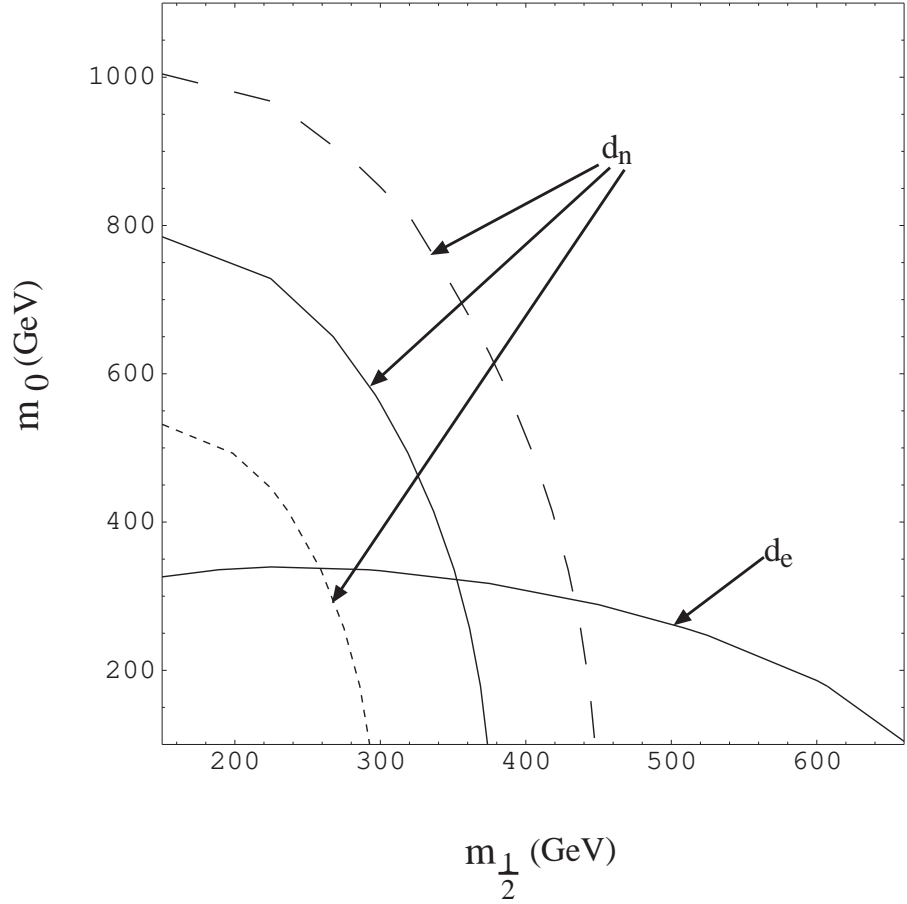
 
\centerline{ \DESepsf(edmfig9.epsf width 12 cm) }
\smallskip
\caption {\label{fig9}The K=0 contours are plotted as a function of $m_0$ and
$m_{1/2}$. The dotted, solid and dashed lines are  for $m_d$ (1 GeV)=5, 8 and 12
MeV respectively. The other input parameters are
$\alpha_{0A}={\pi\over 2}$, $|A_0|= 300$ GeV $\theta_B$=0 and tan$\beta$=3.
Excluded regions are below $d_e$ and to the left of $d_n$ curves.}
\vspace{0 cm}
\end{figure}
\begin{figure}[htb]
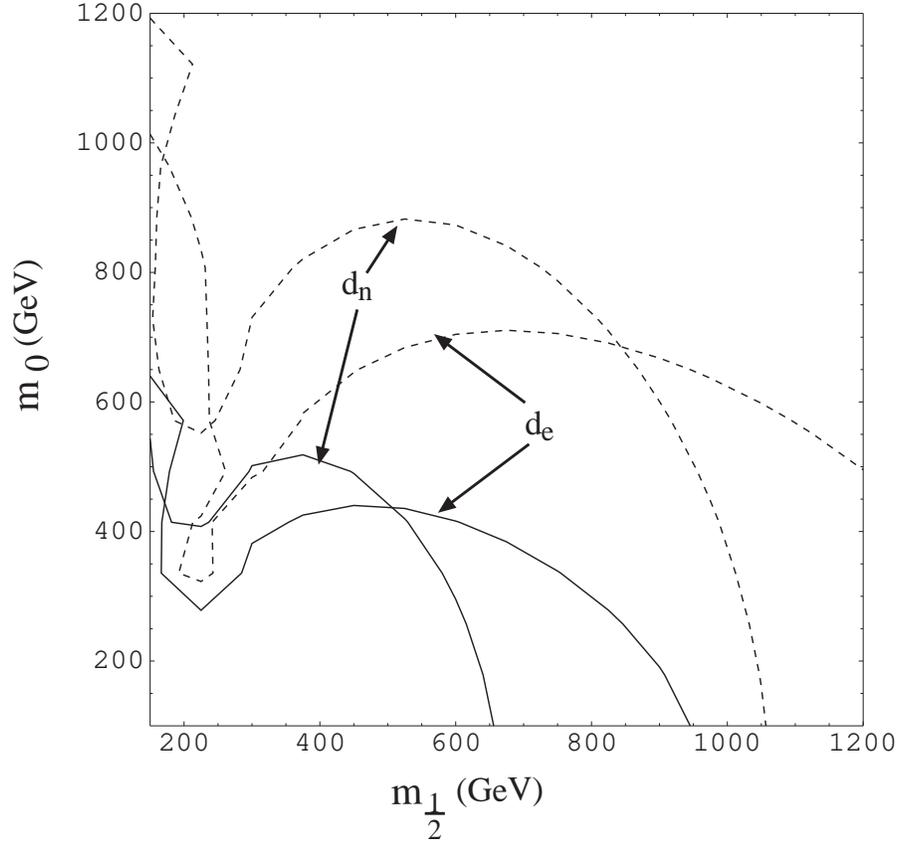
 
\centerline{ \DESepsf(edmfig10.epsf width 12 cm) }
\smallskip
\caption {\label{fig10} Allowed region in $m_0-m_{1/2}$ plane for $d_e$ and
$d_n$.
 The other input parameters are
$\alpha_{0A}={\pi\over 2}$, $|A_0|= 300$ GeV, $\theta_B$=0 and tan$\beta$=20.
 The solid lines are for K=0 and the dotted lines are for K=-0.5.}
\vspace{0 cm}
\end{figure}
\begin{figure}[htb] 
\centerline{ \DESepsf(edmfig250.epsf width 12 cm) }
\smallskip
\caption {\label{fig11}The K=0 contours are plotted as a function of 
$\theta_B$ and $m_{1/2}$, where 
$\alpha_{0A}={\pi\over 2}$, $|A_0|= 300$ GeV, $m_0$=250 GeV and tan$\beta$=3. The
solid lines are for  $d_n$ and the dotted lines are for $d_e$.}
\vspace{0 cm}
\end{figure}
\begin{figure}[htb] 
\centerline{ \DESepsf(edmfig111.epsf width 12 cm) }
\smallskip
\caption {\label{fig12}The K=0 contours are plotted as a function of 
$\theta_B$ and $m_{1/2}$, where 
$\alpha_{0A}={\pi\over 2}$, $|A_0|= 800$ GeV, $m_0$=100 GeV and tan$\beta$=3. The
solid lines are for $d_n$ and the dotted lines are for $d_e$.}
\vspace{0 cm}
\end{figure}
\begin{figure}[htb]
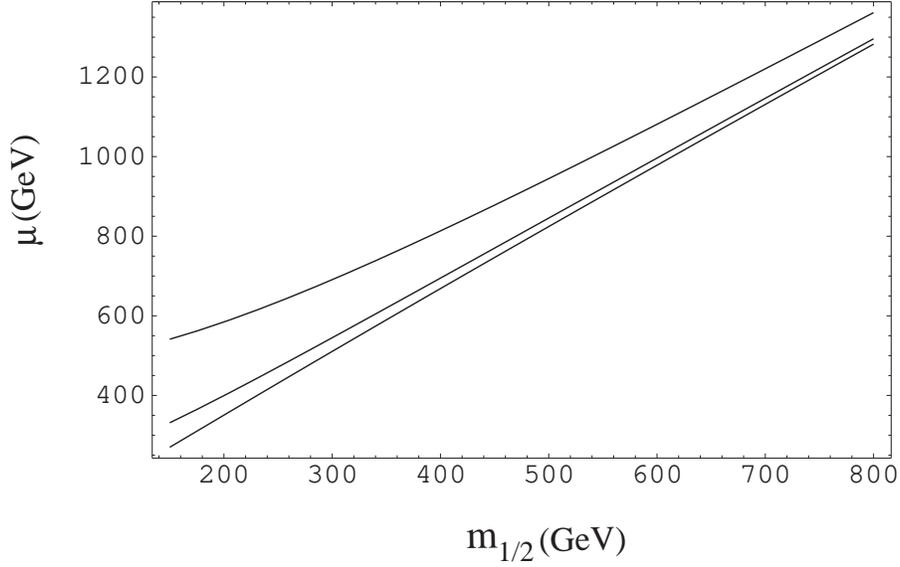
 
\centerline{ \DESepsf(edmfigmyou.epsf width 12 cm) }
\smallskip
\caption {\label{fig13} The parameter $|\mu|$ is plotted as a function of 
$m_{1/2}$. The three solid lines are for $m_0=$ 700 (top), 300 (middle) and 100
(bottom) GeV. The other input parameters are 
$\alpha_{0A}={\pi\over 2}$, $|A_0|= 300$ GeV, tan$\beta$=3 and $\theta_B=0.08$.}
\vspace{0 cm}
\end{figure}

\begin{figure}[htb]
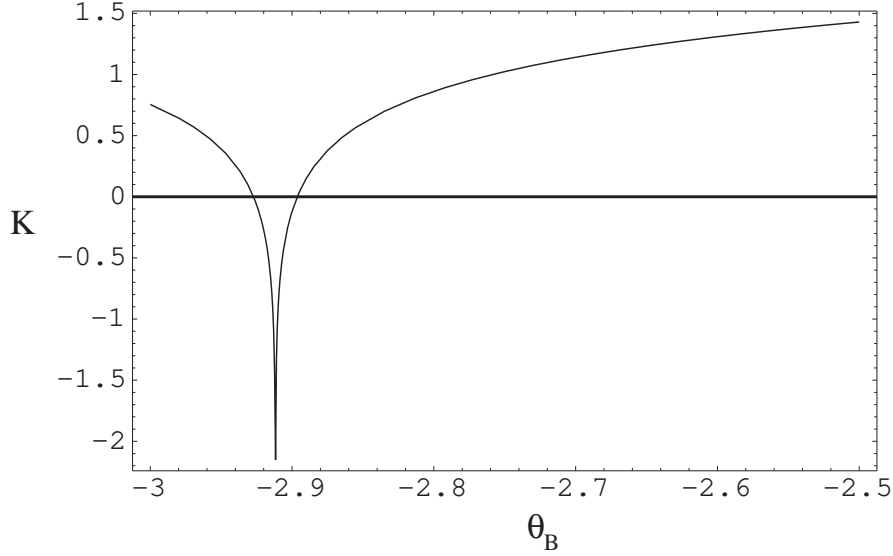
 
\centerline{ \DESepsf(edmnonuni.epsf width 12 cm) }
\smallskip
\caption {\label{fig14} K (for $d_e$)  as a function of $\theta_B$. The other
input parameters are $m_0$=200 GeV; $m_{1/2i}$=150 GeV, tan$\beta$=3,
 $\alpha_{0A}={\pi\over 2}$, $|A_0|= 100$ GeV and $\phi_1=4\pi/5=\phi_3$.}
\vspace{0 cm}
\end{figure}

\begin{figure}[htb]
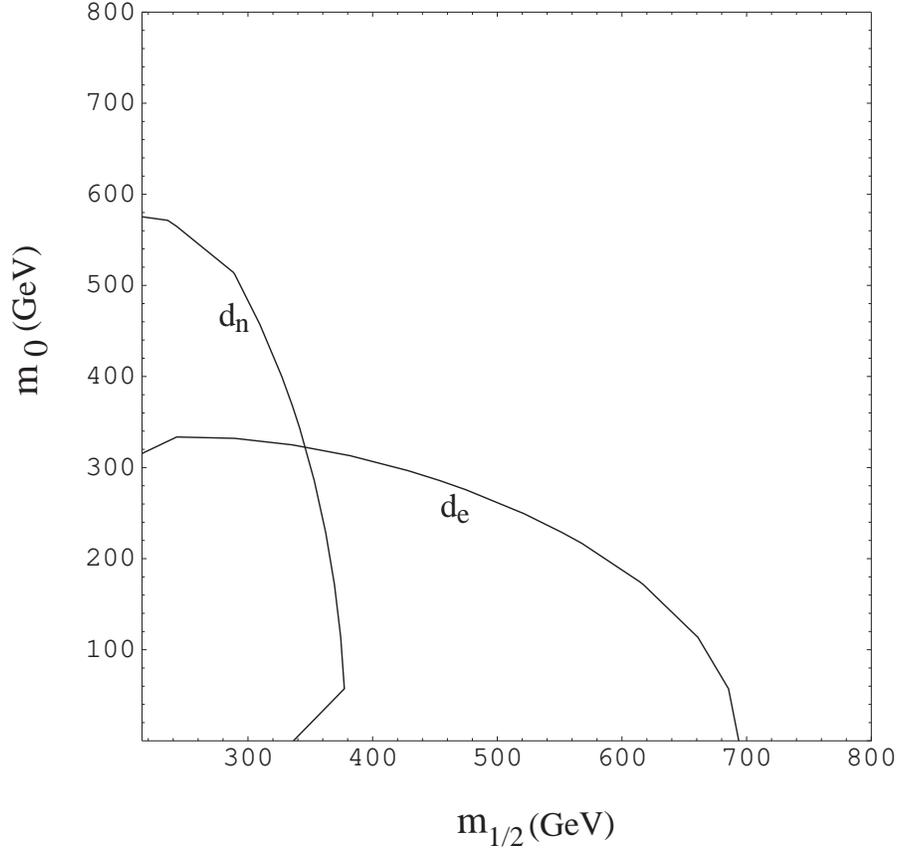
 
\centerline{ \DESepsf(edmfig12.epsf width 12 cm) }
\smallskip
\caption {\label{fig15}Allowed region in $m_0-m_{1/2}$ plane ($K\le0$)
 in a nonuniversal scenario. The other input parameters are
$\alpha_{0A}={\pi\over 2}$, $\theta_B=0$, $|A_0|= 300$ GeV, tan$\beta$=5,
$\delta_1$=1,
$\delta_2$=-1, $\delta_{10}$=-1 and $\delta_{5}$=-0.3.}
\vspace{0 cm}
\end{figure}

\end{document}